\documentclass[floatfix,aps,prl,preprint,superscriptaddress]{revtex4}
\usepackage[dvipdfmx]{graphicx}
\usepackage[]{epsfig}
\usepackage{times,amsmath,amssymb}
\begin{document}

\title{Detection and control of charge states in a quintuple quantum dot}

\author{Takumi Ito}
\email[]{takumi.ito@riken.jp}
\affiliation{Center for Emergent Matter Science, RIKEN, 2-1 Hirosawa, Wako, Saitama 351-0198, Japan}
\affiliation{Department of Applied Physics, University of Tokyo, Bunkyo, Tokyo 113-8656, Japan}

\author{$^{,\dagger} $Tomohiro Otsuka}
\altaffiliation{\ These authors contributed equally to this work.} 
\affiliation{Center for Emergent Matter Science, RIKEN, 2-1 Hirosawa, Wako, Saitama 351-0198, Japan}
\affiliation{Department of Applied Physics, University of Tokyo, Bunkyo, Tokyo 113-8656, Japan}

\author{Shinichi Amaha}
\affiliation{Center for Emergent Matter Science, RIKEN, 2-1 Hirosawa, Wako, Saitama 351-0198, Japan}

\author{Matthieu R. Delbecq}
\affiliation{Center for Emergent Matter Science, RIKEN, 2-1 Hirosawa, Wako, Saitama 351-0198, Japan}
\affiliation{Department of Applied Physics, University of Tokyo, Bunkyo, Tokyo 113-8656, Japan}

\author{Takashi Nakajima}
\affiliation{Center for Emergent Matter Science, RIKEN, 2-1 Hirosawa, Wako, Saitama 351-0198, Japan}
\affiliation{Department of Applied Physics, University of Tokyo, Bunkyo, Tokyo 113-8656, Japan}

\author{Jun Yoneda}
\affiliation{Center for Emergent Matter Science, RIKEN, 2-1 Hirosawa, Wako, Saitama 351-0198, Japan}
\affiliation{Department of Applied Physics, University of Tokyo, Bunkyo, Tokyo 113-8656, Japan}

\author{Kenta Takeda}
\affiliation{Center for Emergent Matter Science, RIKEN, 2-1 Hirosawa, Wako, Saitama 351-0198, Japan}
\affiliation{Department of Applied Physics, University of Tokyo, Bunkyo, Tokyo 113-8656, Japan}

\author{Giles Allison}
\affiliation{Center for Emergent Matter Science, RIKEN, 2-1 Hirosawa, Wako, Saitama 351-0198, Japan}

\author{Akito Noiri}
\affiliation{Center for Emergent Matter Science, RIKEN, 2-1 Hirosawa, Wako, Saitama 351-0198, Japan}
\affiliation{Department of Applied Physics, University of Tokyo, Bunkyo, Tokyo 113-8656, Japan}

\author{Kento Kawasaki}
\affiliation{Center for Emergent Matter Science, RIKEN, 2-1 Hirosawa, Wako, Saitama 351-0198, Japan}
\affiliation{Department of Applied Physics, University of Tokyo, Bunkyo, Tokyo 113-8656, Japan}

\author{Seigo Tarucha}%
\affiliation{Center for Emergent Matter Science, RIKEN, 2-1 Hirosawa, Wako, Saitama 351-0198, Japan}
\affiliation{Department of Applied Physics, University of Tokyo, Bunkyo, Tokyo 113-8656, Japan}
\affiliation{Quantum-Phase Electronics Center, University of Tokyo, Bunkyo, Tokyo 113-8656, Japan}
\affiliation{Institute for Nano Quantum Information Electronics, University of Tokyo, 4-6-1 Komaba, Meguro, Tokyo 153-8505, Japan}

\date{\today}
\begin{abstract}
A semiconductor quintuple quantum dot with two charge sensors and an additional contact to the center dot from an electron reservoir is fabricated to demonstrate the concept of scalable architecture. 
This design enables formation of the five dots as confirmed by measurements of the charge states of the three nearest dots to the respective charge sensor. 
The gate performance of the measured stability diagram is well reproduced by a capacitance model.
These results provide an important step towards realizing controllable large scale multiple quantum dot systems.
\end{abstract}

\maketitle

%%% Introduction %%% 

%quantum computing, multi spin physics%
Quantum dots (QDs) are artificial systems in which electrons are confined in all three dimensions and the electronic states are determined by the confining potential and Coulombic interaction~\cite{1997Kouwenhovenreview}. 
For multiple QDs the electronic states are furthermore influenced by the tunneling and interaction between dots. 
QDs can offer intriguing systems for constructing fermion Hubbard models~\cite{2008ByrnesPRB} and also implementing elements of quantum computing~\cite{1995DiVicenzoScience, 1998LossPRA, 2005TaylorNatPhys, 2007HansonRMP}. 
Increasing the number of QDs is a necessary step towards these goals and has been attempted using various kinds of materials such as semiconductor heterostructures, nanowires~\cite{2004BjorkNanoLett, 2005FasthNanoLett} and self-assembled dots~\cite{1996CusackPRB, 1998GarciaAPL, Fonseca1998PRB}. 
Single to quadruple QDs have been fabricated in semiconductor heterostrcuture~\cite{2012ThalineauAPL, 2014TakakuraAPL, 2014DelbecqAPL} and applied to quantum bits using the charge or spin degree of freedom~\cite{2005PettaScience, 2006KoppensNature, 2010TakakuraAPL, 2013MedfordPRL, 2012ShulmanScience, 2014YonedaPRL, 2015OtsukaarXiv}. 

\begin{figure}
\begin{center}
  \includegraphics[scale=0.25]{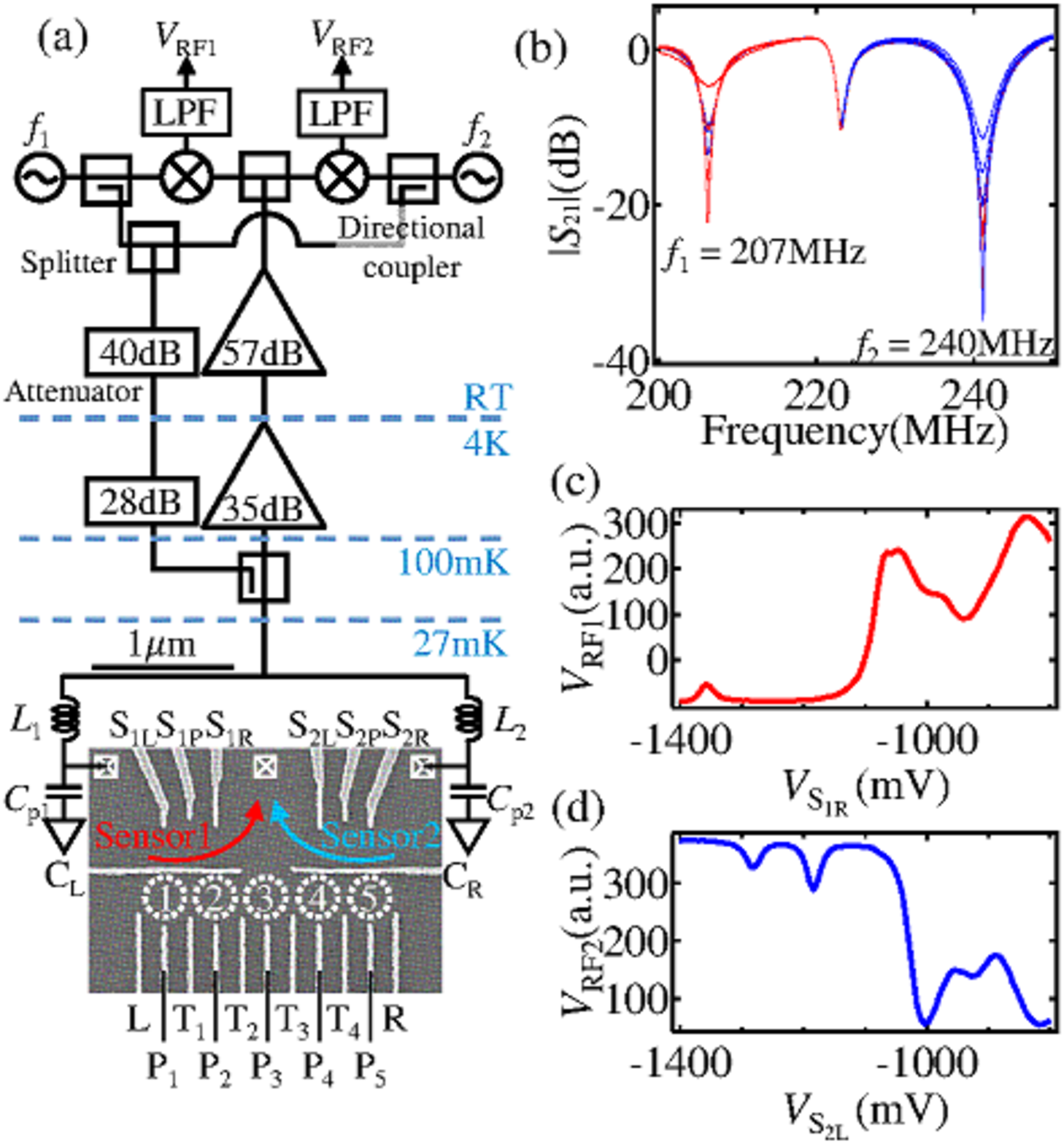}
  \caption{(color online)
(a) Scanning electron micrograph of the 5QD device and schematic of the measurement setup. 
(b) $|S_{21}|$ of the resonance circuit as a function of the carrier microwave frequency. The left (right) dip is caused by the resonator including sensor 1 (sensor 2).
  The center dip is caused by an unused resonator not connected to the device.
  The traces show the results with different conductance of the sensors (sensor 1 from 0.88 to 0.19 $e^{2}/h$ and sensor 2 from 0.77 to 0.03 $e^{2}/h$).
  (c) ((d)) Changes of the RF signal from sensor1, $V_{\rm RF1}$, as a function of $V_{\rm S_{\rm 1R}}$ (c) and from sensor2, $V_{\rm RF2}$ as a function of $V_{\rm S_{\rm 2L}}$ (d).}
\label{Device}
\end{center}
\end{figure}

Scale-up of QD systems whose electronic states can be precisely manipulated and detected requires several technical advances. 
In the conventional device architecture, the electronic states are electrically manipulated by two plunger gates and detected by a single charge sensor~\cite{1993FieldPRL, 1998SchoelkopfScience, 2007ReillyAPL, 2007CassidyAPL, 2007MullerAIP, 2010BarthelPRB}.  
Double or triple QDs (DQD or TQD) are the typical cases in which the charge states can be manipulated by two plunger gates attached to the two dots and detected by a charge sensor.
This technique has been applied to quadruple QDs but not more, probably because the sensor sensitivity decreases with the distance to the target QD and also because more plunger gates must be appropriately adjusted to address the individual QDs. 
In addition multiple QDs are usually constructed by connecting dots in a row with a tunnel-coupled reservoir at each end. 
This geometry makes it difficult to load electrons from the reservoirs to the inner dots ~\cite{2009GaudreauAPL}. 
In general a set of two plunger gates, one charge sensor and two reservoirs is appropriate to address a triple QD.
Therefore splitting into TQDs may be a straightforward approach to scale up the QD architecture ~\cite{2010LairdPRB, 2014HornibrookAPL}. 

In this work, we fabricate a semiconductor quintuple quantum dot (QuiQD) or series coupled five QDs with a concept relevant for further increasing the number of QDs. 
Our QuiQD has a reservoir connected to the leftmost, center and rightmost dots, to facilitate loading of electrons to all dots. 
In addition, two RF charge sensors are independently and simultaneously operated using a frequency multiplexing technique~\cite{2010LairdPRB} to complementarily and precisely read out the charge states. 
We modify the charge configuration with gate voltages to demonstrate the utility of the new architecture by comparing the measured stability diagrams with capacitance model calculations.

%%% Device & measurement setup %%%

%device%
Figure~\ref{Device}(a) shows a scanning electron micrograph of the device and a schematic of the measurement setup. 
The device was fabricated from a GaAs/AlGaAs heterostructure wafer with an electron sheet carrier density of 5.6~$\times$~10$^{15}$~m$^{-2}$ and a mobility of 17$ $~m$^2$/Vs. 
The two-dimensional electron gas is formed 60~nm under the wafer surface. We patterned a mesa by wet-etching and formed Ti/Au Schottky surface gates by metal deposition, which appear white in Figure~\ref{Device}(a). 
By applying negative voltages to the gate electrodes, five QDs (QD$_{1}$ to QD$_{5}$), and two QD charge sensors (sensors 1 and 2) are formed at the dotted circles, and arrows, respectively. 
Sensors 1 and 2 can efficiently detect the three leftmost dots (QD$_{1}$ to QD$_{3}$), and the three rightmost (QD$_{3}$ to QD$_{5}$) dots, respectively. 
The plunger gate P$_{i}$ tunes predominantly the energy level of QD$_{i}$, while the tunnel gate T$_{i}$ tunes the tunnel coupling between QD$_{i}$ and QD$_{i+1}$. 
To induce an additional reservoir coupling at QD$_{3}$, a gap is made in the horizontal line gate (between C$_{\rm L}$ and C$_{\rm R}$). 
Electrons are then loaded from the two reservoirs to all dots. 
This helps to initialize the charge states of the QuiQD. 
All measurements were conducted in a dilution fridge cryostat with a base temperature of 27mK.

%measurement%
\begin{figure}
\begin{center}
  \includegraphics[scale=0.25]{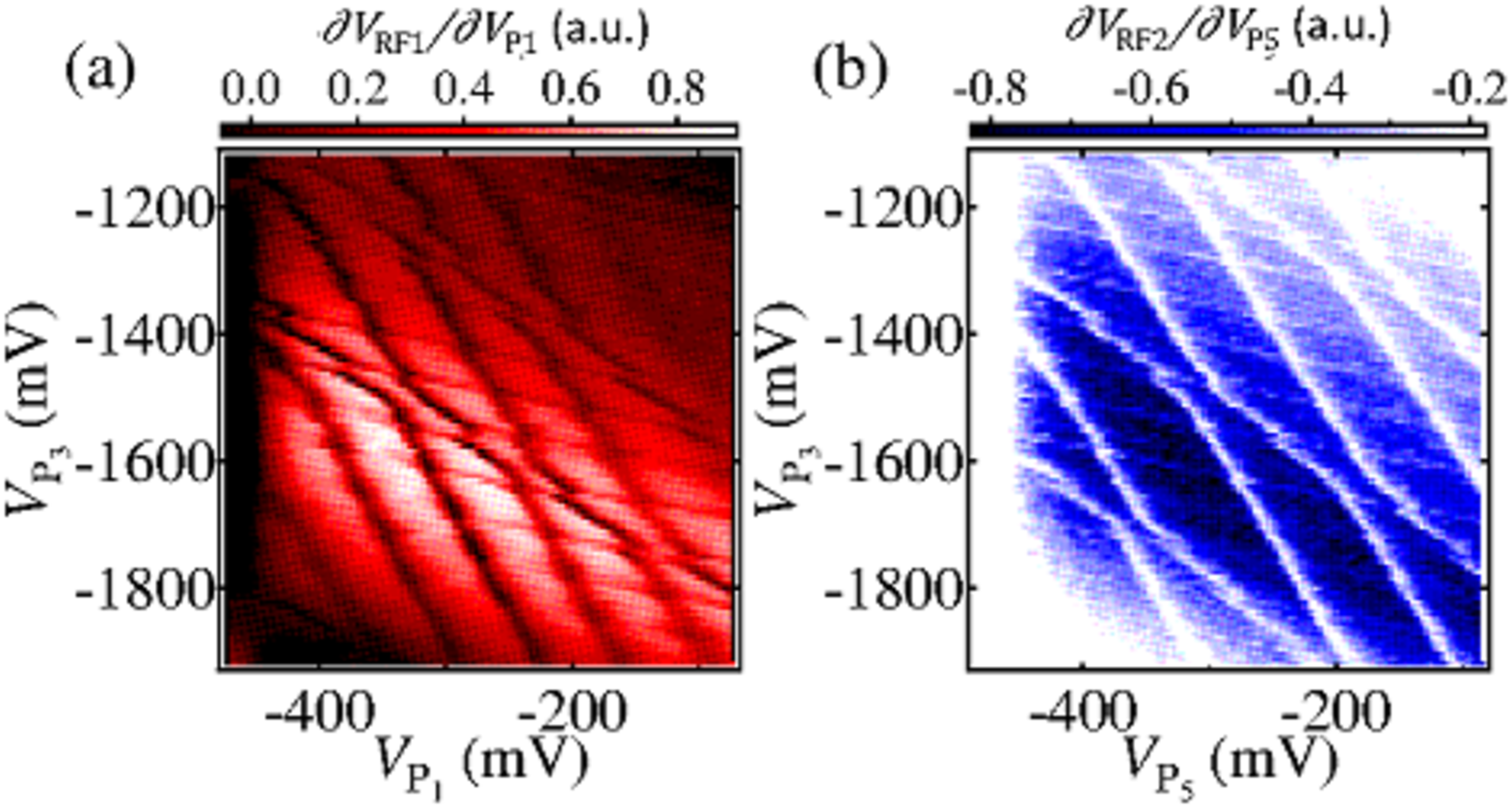}
  \caption{(color online) 
  (a) ((b)) (a)	Stability diagram measured for QD$_{1}$ to QD$_{3}$ in $\partial V_{\rm RF1}/\partial V_{\rm P_{1}}$ as a function of $V_{\rm P_{1}}$ and $V_{\rm P_{3}}$ (a) and for QD$_{3}$ to QD$_{5}$ in $\partial V_{\rm RF2}/\partial V_{\rm P_{5}}$ as a function of $V_{\rm P_{5}}$ and $V_{\rm P_{3}}$ (b). 
The QuiQD can be tuned by separating it into two TQDs since QD$_3$ is contacted to the lead.
}
  \label{P1-P3}
\end{center}
\end{figure}

The QD charge sensors are connected to RF resonators configured by the inductors $L_{\rm 1}$ =270 nH and $L_{\rm 2}$ =470 nH and the stray capacitances $C_{\rm p1}$ and $C_{\rm p2} (\approx 0.4$pF) for the RF reflectometry. 
Figure~\ref{Device}(b) shows the reflected RF signal $|S_{21}|$ from the resonance circuit measured by the setup of Figure~\ref{Device}(a). 
We observe dips caused by the resonance circuits including sensor 1 and sensor 2 at 207 MHz and 240 MHz respectively. 
The red (blue) trace shows the conductance of sensor 1 (sensor 2). 
We can detect the change of the sensor conductance through the reflected signal: $|S_{21}|$ at $ f_{1}$ changes by 17dB due to the conductance change of sensor 1 from 0.88 to 0.19 $e^{2}/h$. 
Similarly the reflected signal at $ f_{2}$ changes by 23dB depending on the conductance change of sensor 2 from 0.77 to 0.03 $e^{2}/h$.

To read out the reflected signals at different frequencies, the room temperature part of the measurement circuit is configured by two sets of local oscillators and mixers (Figure~\ref{Device}(a)). 
In this room temperature circuit, two RF carriers are combined and the reflected signal of each charge sensor is picked up by the mixer operating at each carrier frequency simultaneously. 
Note that simultaneous readout may be important for measurement of temporal correlation of charge or spin between different dots~\cite{2013BraakmanNatNanotech, 2015SrinivasaPRL}. 
The changes of the RF signal from sensor 1 ($V_{\rm RF1}$) and sensor 2 ($V_{\rm RF2}$) are shown in Figures~\ref{Device}(c) and (d) as a function of the sensor gate voltages $V_{\rm S_{1R}}$ and $V_{\rm S_{2R}}$ respectively. 
Note that due to a difference in phase, the reflected signals change in opposite directions in Figure~\ref{Device}(d). 
In the following measurement, gate voltages $V_{\rm S_{1R}}$ and $V_{\rm S_{2R}}$ are adjusted to the condition most sensitive to electrostatic changes of the surrounding environment.

%%% Tuning of 5QD %%%
\begin{figure}
\begin{center}
  \includegraphics[scale=0.25]{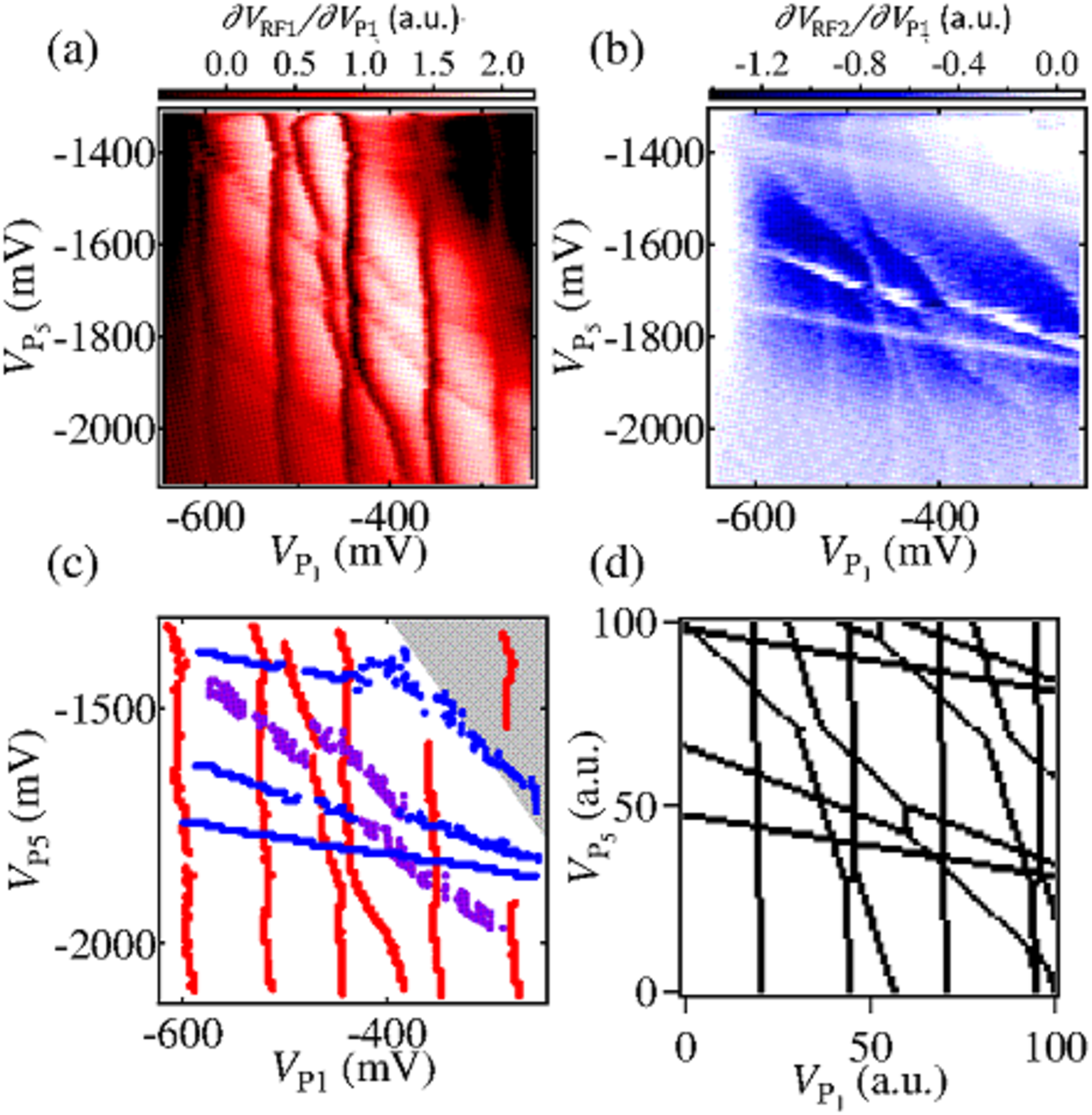}
  \caption{(color online) Stability diagram in the plane of $V_{\rm P_1}$ and $V_{\rm P_5}$ for the QuiQD measured simultaneously using the multiplex technique: $\partial V_{\rm RF1}/\partial V_{\rm P_{1}}$ (a) and $\partial V_{\rm RF1}/\partial V_{\rm P_{1}}$ (b) with $V_{\rm P_2}$=-1585 mV, $V_{\rm P_3}$=-1020 mV, and $V_{\rm P_4}$=-470mV. 
  (c) Data points extracted from the charge transition lines in (a) and (b): Red, or blue points from (a), or (b), respectively. 
  The grey region shows the area where the sensor sensitivity is too low to apparently distinguish the transition lines. 
  (d) Calculated stability diagram using the capacitive QD model. The capacitance values are estimated from the experiments..}
  \label{5QD}
\end{center}
\end{figure}

%3+3 tuning%
Gate tuning of the QuiQD is simplified by splitting the five QDs into two TQDs and manipulating the charge states on the two different stability diagrams. 
Figure~\ref{P1-P3} shows the numerical derivative of the RF reflectometry signal measured by sensor 1, ${\rm \partial}V_{\rm RF1}/{\rm \partial}V_{\rm P_{1}}$, in the $V_{\rm P_{1}}$- $V_{\rm P_{3}}$ plane (a) and by sensor 2, ${\rm \partial}V_{\rm RF2}/{\rm \partial}V_{\rm P_{5}}$, in the $V_{\rm P_{5}}$ - $V_{\rm P_{3}}$ plane (b), respectively. 
In each diagram, we observe three sets of distinct charge transition lines with three different slopes, which are defined by the capacitive couplings between the dots and the modulating gates. 
Each set of charging lines (from the more horizontal to the more vertical) is assigned to charging QD$_{1}$ to QD$_{3}$ in (a) and QD$_{5}$ to QD$_{3}$ in (b). 
We adjust the voltages on $V_{\rm T_{1}}$, $V_{\rm T_{2}}$, $V_{\rm T_{3}}$ and $V_{\rm T_{4}}$ to make all tunnel or electrostatic couplings between adjacent dots roughly the same judging from the size of avoided crossings between two different charge transition lines. 
Here we confirm that there are no apparent couplings present between distant dots, because the corresponding charging lines just cross with each other with no anticrossing. 
%%From the two diagrams we evaluate the gate capacitances of P$_{1}$, P$_{3}$, and P$_{5}$ to the directly and indirectly connected dots QD$_{1}$, QD$_{3}$, and QD$_{5}$. 
Since the two diagrams share a common P$_{3}$ axis in the same range, we are able to evaluate appropriate voltages of all gates to manipulate the charge state of the QuiQD.

%%%5QD stability diagram%%%

%5QD detection%
We use the gate voltage setting derived from Figure~\ref{P1-P3} as a guide to establish the stability diagram of the QuiQD. 
Figures~\ref{5QD} (a) and (b) show the diagram in the plane of $V_{\rm P_1}$ and $V_{\rm P_5}$ measured using sensor 1 ($\partial V_{\rm RF1}/\partial V_{\rm P_{1}}$) and 2 ($\partial V_{\rm RF2}/\partial V_{\rm P_{1}}$), respectively. 
The other gate voltages are fixed at $V_{\rm P_2}$=-1585mV, $V_{\rm P_3}$=-1020mV, and $V_{\rm P_4}$=-470mV. 
The values of $V_{\rm T_1}$ to $V_{\rm T_4}$ are the same as used in Figure~\ref{P1-P3}. 
In both figures, five sets of charge transition lines with different slopes are distinguished and from the slopes we are able to assign them to charging five different dots: QD$_{1}$ to QD$_{5}$ from vertical to horizontal. 
Figures~\ref{5QD}(a) and (b) are measured simultaneously using the multiplex technique of RF reflectometry. 
Note the charge transition lines of QD$_{1}$ to QD$_{3}$ are clearly visible whereas those of QD$_{4}$ and QD$_{5}$ are less visible in Figure ~\ref{5QD} (a). 
In contrast the charging lines of QD$_{3}$ to QD$_{5}$ are more visible in Figure~\ref{5QD}(b). 
This observation indicates that each sensor is sensitive to charging of at least three nearest QDs and that two sensors can together detect all charge transitions of the QuiQD.
Note that the dots in Figures~\ref{5QD} (a) and (b) are not in a few electron regime due to limitation of the gate voltage range and contain dozens of electrons judging from the spacing of the charge transition lines~\cite{2014TakakuraAPL}. 
Also QD$_3$ has the most electrons due to the gate electrode design. 
We will be able to reduce the number of electrons by reducing the gaps between the gates to form smaller dots.

In Figure~\ref{5QD} (c) we show the charging lines for the QuiQD by plotting the data points of the dark and white lines in Figures~\ref{5QD} (a) and (b): red and blue points from (a) and (b) and purple points from both. 
Avoided crossings of charging lines of neighboring QDs indicate finite capacitive coupling among all five QDs as is the case in Figure~\ref{P1-P3}. 
Also, none of the charge transition lines are fragmented, suggesting that tunneling rates are kept sufficiently high for all QDs. 
Note that charge sensors are tuned to be most sensitive at the center of stability diagrams and become insensitive in the upper right region (grey region of Figure~\ref{5QD}(c)).

In large systems of multiple QDs, the charge states become complicated and difficult to discriminate. 
Therefore numerical calculations of stability diagrams are helpful in the process of adjusting gate voltages to search for desirable charge states. 
We find that the charge stability diagram obtained here is well reproduced in a qualitative manner using a capacitive QD model~\cite{2014DelbecqAPL, 2003WielRMP}. 
Figure~\ref{5QD} (d) is the calculated stability diagram to reproduce the experiment of Figure~\ref{5QD} (c). 
The ratios of the capacitance used in the calculation are all taken from the experiment.
This simple model shows good agreement with the experiment in which the dots contain many electrons and when we focus on a limited range of the charge stability diagram.
We see that the main features in Figure~\ref{5QD} (c) are well reproduced by the calculation.

%Tunability
\begin{figure}
\begin{center}
  \includegraphics[scale=0.25]{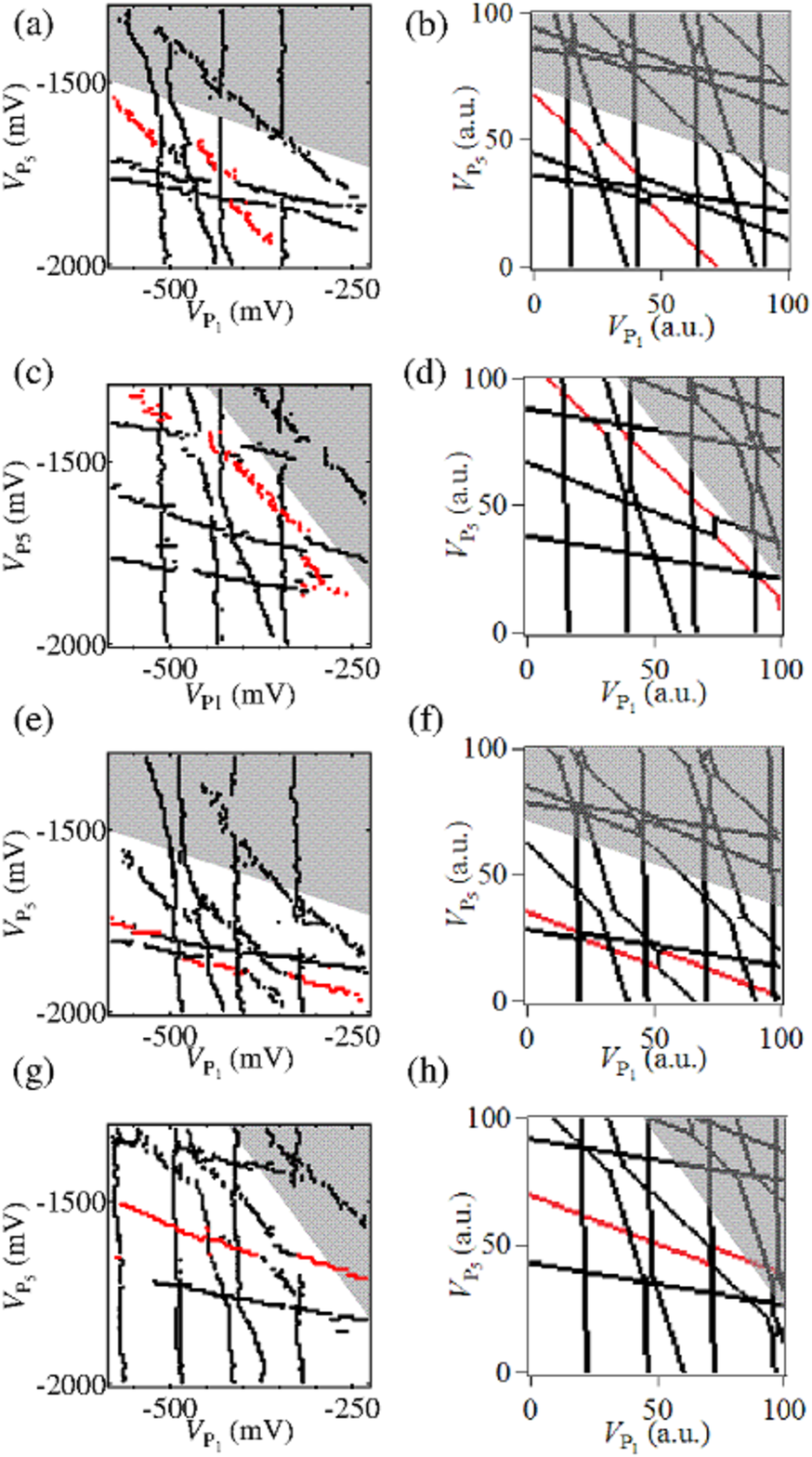}
  \caption{(color online) Comparison between the measured and calculated stability diagrams in the plane of $V_{\rm P_{1}}$ and $V_{\rm P_{5}}$ with $V_{\rm P_{3}}$ and $V_{\rm P_{4}}$ as parameters: $V_{\rm P_{3}}$ = − 1000 mV and $V_{\rm P_{4}}$= − 470mV in (a) and (b);
 $V_{\rm P_{3}}$= − 1040 mV and $V_{\rm P_{4}}$= − 470 mV in (c) and (d); 
 $V_{\rm P_{3}}$= − 1020 mV and $V_{\rm P_{4}}$= − 450 mV in (e) and (f); 
 $V_{\rm P_{3}}$= − 1020 mV and $V_{\rm P_{4}}$= − 490 mV in (g) and (h). 
 The grey shows the area where some charging lines are not distinguished due to the low sensor sensitivity.}
  \label{modulate}
\end{center}
\end{figure}

Finally we demonstrate the tunability of this device. Figures~\ref{modulate}(a) and (c) are the $V_{\rm P_1}$- $V_{\rm P_5}$ stability diagrams measured for two different $V_{\rm P_3}$ values of -1000 mV, and -1040 mV, respectively but keeping other gate voltage values the same as in Figure~\ref{5QD}(c).
 The charge transition line of QD$_{3}$ highlighted in red shifts more than the other charge transition lines. 
This shift is well reproduced by the calculation of Figure~\ref{modulate}(b) an (d). 
In the same way, Figures~\ref{modulate}(e), and (g) are the diagrams measured for two different values of $V_{\rm P_4}$ of -450 mV and -490mV, respectively. 
The charging line of QD$_4$ highlighted in red shifts more than the others as expected from the calculation of Figures~\ref{modulate} (f) and (h).

%%Conclusion%% 
In conclusion, we have fabricated a QuiQD device with an additional contact to the center dot from a reservoir and two RF charge sensors, whose design suits further increasing of the number of QDs. 
We have characterized the gate performance on the charge state stability diagram and well distinguished the charge transition lines corresponding to all five dots thanks to the use of the two charge sensors. 
We have demonstrated that the gate performance on the stability diagram is well reproduced by the capacitance model. 
These results are important steps for further scale up of QD system.

%%Acknowledgement%%
We thank J. Beil, J. Medford, F. Kuemmeth, C. M. Marcus, D. J. Reilly, K. Ono, RIKEN CEMS Emergent Matter Science Research Support Team and Microwave Research Group in Caltech for fruitful discussions and technical supports.
Part of this work is supported by Funding Program for World-Leading Innovative R\&D on Science and Technology (FIRST), 
the Grant-in-Aid for Research Young Scientists B from the Japan Society for the Promotion of Science, 
ImPACT Program of Council for Science, 
Technology and Innovation, 
CREST, 
JST, 
Yazaki Memorial Foundation for Science and Technology Research Grant, 
Japan Prize Foundation Research Grant, 
Advanced Technology Institute Research Grant, 
the Murata Science Foundation Research Grant, 
Izumi Science and Technology Foundation Research Grant, 
IARPA project``Multi-Qubit Coherent Operations`` through Copenhagen University.


\begin{references}

\bibitem{1997Kouwenhovenreview}
L. P. Kouwenhoven, C. M. Marcus, P. L. McEuen, S. Tarucha, R. M. Westervelt, and N. S. Wingreen, Proceedings of the NATO Advanced Study Institute on Mesoscopic Electron Transport, 105, (1997)

\bibitem{2008ByrnesPRB}
T. Byrnes, N. Y. Kim, K. Kusudo, and Y. Yamamoto, Phys. Rev. B \textbf{78}, 075320 (2008).

\bibitem{1995DiVicenzoScience}
D. P. DiVincenzo, Science \textbf{270}, 255 (1995).


\bibitem{1998LossPRA}
D. Loss and D. P. DiVincenzo, Phys. Rev. A \textbf{57}, 120 (1998).

\bibitem{2005TaylorNatPhys}
J. M. Taylor, H. -A. Engel, W. D\"{u}r, A. Yacoby, C. M. Marcus, P. Zoller, and M. D. Lukin, Nat. Phys.\textbf{1}, 177 (2005).

\bibitem{2007HansonRMP}
R. Hanson, L. P. Kouwenhoven, J. R. Petta, S. Tarucha, and L. M. K. Vandersypen, Rev. Mod. Phys. \textbf{79}, 1217 (2007).

\bibitem{2004BjorkNanoLett}
M. T. Bj\"{o}rk, C. Thelander, A. E. Hansen, L. E. Jensen, M. W. Larsson, L. R. Wellenberg, and L. Samuelson, Nano Lett. \textbf{4}, 1621 (2004).

\bibitem{2005FasthNanoLett}
C. Fasth, A. Fuhrer, M.T. Bj\"{o}rk, and L. Samuelson, Nano Lett. \textbf{5}, 1487 (2005).

\bibitem{1996CusackPRB}
M. A. Cusack, P. Briddon, and M. Jaros, Phys. Rev. B. \textbf{54}, R2300 (1996).

\bibitem{1998GarciaAPL}
J. M. Garc\'{i}a, T. Mankad, P. O. Holtz, P. J. Wellman, and P. M. Petroff, Appl. Phys. Lett. \textbf{72}, 3172 (1998).

\bibitem{Fonseca1998PRB}
L. Fonseca, J. Jimenez, and J. Leburton, Phys. Rev. B \textbf{58}, 9955 (1998).


\bibitem{2012ThalineauAPL}
R. Thalineau, S. Hermelin, A. D. Wieck, C. B\"{a}uerle, L. Saminadayar, and T. Meunier, Appl. Phys. Lett. \textbf{101}, 103102 (2012).

\bibitem{2014TakakuraAPL}
T. Takakura, A. Noiri, T. Obata, T. Otsuka, J. Yoneda, K. Yoshida, and S. Tarucha, Appl. Phys. Lett. \textbf{104}, 113109 (2014).

\bibitem{2014DelbecqAPL}
M. R. Delbecq, T. Nakajima, T. Otsuka, S. Amaha, J. D. Watson, M. J. Manfra, and S. Tarucha, Appl. Phys. Lett. \textbf{104}, 183111 (2014).

\bibitem{2005PettaScience}
J. R. Petta, A. C. Johnson, J. M. Taylor, E. A. Laird, A. Yacoby, M. D. Lukin, C. M. Marcus, M. P. Hanson, and A. C. Gossard, Science \textbf{309}, 2180 (2005).

\bibitem{2006KoppensNature}
F. H. L. Koppens, C. Buizert, K. J. Tielrooij, I. T. Vink, K. C. Nowack, T. Meunier, L. P. Kouwenhoven, and L. M. K. Vandersypen, Nature \textbf{442}, 766 (2006).

\bibitem{2010TakakuraAPL}
T. Takakura, M. Pioro-Ladri\'{e}re, T. Obata, Y. S. Shin, R. Brunner, K. Yoshida, T. Taniyama, and S. Tarucha, Appl. Phys. Lett. \textbf{97}, 212104 (2010).

\bibitem{2013MedfordPRL}
J. Medford, J. Beil, J. M. Taylor, E. I. Rashba, H. Lu, A. C. Gossard, and C. M. Marcus, Phys. Rev. Lett.\textbf{ 111}, 050501 (2013).

\bibitem{2012ShulmanScience}
M. D. Shulman, O. E. Dial, S. P. Harvey, H. Bluhm, V. Umansky, and A. Yacoby, Science \textbf{336}, 202 (2012).

\bibitem{2014YonedaPRL}
J. Yoneda, T. Otsuka, T. Nakajima, T. Takakura, T. Obata, M. Pioro-Ladri\'{e}re, H. Lu, C. J. Palmstr\o m, A. C. Gossard, and S. Tarucha, Phys. Rev. Lett. \textbf{113}, 267601 (2014).

\bibitem{2015OtsukaarXiv}
T. Otsuka, T. Nakajima, M. R. Delbecq, S. Amaha, J. Yoneda, K. Takeda, G. Allison, T. Ito, R. Sugawara, A. Noiri, A. Ludwig, A. D. Wieck, and S. Tarucha, arXiv:1510.02547  (2015).

\bibitem{1993FieldPRL}
M. Field, C. Smith, M. Pepper, D. Ritchie, J. Frost, G. Jones, and D. Hasko, Phys. Rev. Lett. \textbf{70}, 1311 (1993).

\bibitem{1998SchoelkopfScience}
R. J. Schoelkopf, P. Wahlgren, A. A. Kozhevnikov, P. Deising, and D. E. Prober,  Science \textbf{280}, 1238 (1998).

\bibitem{2007ReillyAPL}
D. J. Reilly, C. M. Marcus, M. P. Hanson, and A. C. Gossard, Appl. Phys. Lett. \textbf{91}, 162101 (2007).

\bibitem{2007CassidyAPL}
M. C. Cassidy, A. S. Dzurak, R. G. Clark, K. D. Petersson, I. Farrer, D. A. Ritchie, and C. G. Smith, Appl. Phys. Lett. \textbf{91}, 222104 (2007).

\bibitem{2007MullerAIP}
T. M\"{u}ler, K. Vollenweider, T. Ihn, R. Schleser, M. Sigrist, K. Ensslin, M. Reinwald, and W. Wegscheider, AIP Conf. Proc. \textbf{893}, 1113 (2007).

\bibitem{2010BarthelPRB}
C. Barthel, M. Kj\ae rgaard, J. Medford, M. Stopa, C. M. Marcus, M. P. Hanson, and a. C. Gossard, Phys. Rev. B \textbf{81}, 161308 (2010).

\bibitem{2009GaudreauAPL}
L. Gaudreau, A. Kam, G. Granger, S. A. Studenikin, P. Zawadzki, and A. S. Sachrajda, Appl. Phys. Lett. \textbf{95}, 193101 (2009).

\bibitem{2010LairdPRB}
E. A. Laird, J. M. Taylor, D. P. Divincenzo, C. M. Marcus, M. P. Hanson, and A. C. Gossard, Phys. Rev. B \textbf{82}, 075403 (2010).

\bibitem{2014HornibrookAPL}
J. M. Hornibrook, J. I. Colless, A. C. Mahoney, X. G. Croot, S. Blanvillain, H. Lu, A. C. Gossard, and D. J. Reilly, Appl. Phys. Lett. \textbf{104}, 103108 (2014).

\bibitem{2013BraakmanNatNanotech}
F. Braakman, P. Barthelemy, C. Reichl, W. Wegschneider, and L. M. K. Vandersypen, Nat. Nanotechnol. \textbf{8}, 432-437 (2013).

\bibitem{2015SrinivasaPRL}
V. Srinivasa, H. Xu, and J. M. Taylor, Phys. Rev. Lett. \textbf{114}, 226803 (2015).

\bibitem{2003WielRMP}
W. G. Van Der Wiel, S. De Franceschi, J. M. Elzerman, T. Fujisawa, S. Tarucha, and L. P. Kouwenhovens Rev. Mod. Phys. \textbf{75}, 1 (2003).



\end{references}
\end{document}